\pdfoutput=1 
﻿

\documentclass[aps,prd,reprint,preprintnumbers,superscriptaddress,showpacs,twocolumn]{revtex4-1}
\usepackage{latexsym,graphicx,amssymb,amsmath,mathrsfs}
\usepackage{setspace,bm}
\usepackage[breaklinks, colorlinks=true, pdfstartview=FitV, linkcolor=red, citecolor=blue, urlcolor=blue]{hyperref}
\usepackage[usenames]{color}
\usepackage{latexsym}
\usepackage{epstopdf}
\usepackage{mathtools}
\usepackage{url}
\usepackage{comment}
\usepackage{braket}

\allowdisplaybreaks[1]
\usepackage[normalem]{ulem}

\renewcommand\sout{\bgroup \color{red} \ULdepth=-.5ex \ULset}

\begin{document}

\title{
Anatomy of the dense QCD matter from canonical sectors
}

\author{Kouji Kashiwa}
\email[]{kashiwa@fit.ac.jp}
\affiliation{Fukuoka Institute of Technology, Wajiro, Fukuoka 811-0295,
Japan}

\author{Hiroaki Kouno}
\email[]{kounoh@cc.saga-u.ac.jp}
\affiliation{Department of Physics, Saga University, Saga 840-8502,
Japan}

\begin{abstract}
We investigate the nuclear and the quark matter at finite real chemical potential ($\mu_\mathrm{R}$) and low temperature from the viewpoint of the canonical sectors constructed via the imaginary chemical potential region.
Based on the large $N_\mathrm{c}$ estimation, where $N_\mathrm{c}$ is the number of color, we can discuss the confinement-deconfinement nature at finite $\mu_\mathrm{R}$ from the canonical sectors.
We found the expectation that the sharp change of canonical sectors at $\mu_\mathrm{R} \sim M_\mathrm{B}/N_\mathrm{c}$, where $M_\mathrm{B}$ is the lowest baryon mass, is happen in the large $N_\mathrm{c}$ regime, and it is matched with the quarkyonic picture.
In addition, we discussed the color superconductivity and the chiral properties from the structure of canonical sectors.
Even in the present anatomy from the canonical sectors, we can have the suitable picture for the dense QCD matter.
 \end{abstract}
\maketitle

\section{Introduction}

Exploring the phase structure of Quantum Chromodynamics (QCD) at finite temperature ($T$) and real chemical potential ($\mu_\mathrm{R}$) is an important and interesting subject in the several research fields such as elementary particle, hadron and nuclear physics.
Particularly, the moderate $\mu_\mathrm{R}$ region attracts much more attention recently because some exotic phases such as the color-superconducting, quarkyonic and inhomogeneous chiral symmetry broken phases are expected to be appeared in the region; for example, see Ref.\,\cite{Fukushima:2010bq}.
If we can obtain the QCD phase diagram starting from the first principle calculation that is the lattice QCD simulation, there are no unclearness, but it is not feasible at moderate $\mu_\mathrm{R}$ because of the well known sign problem; see Ref.\,\cite{deForcrand:2010ys} for a review of the sign problem and Refs.\,\cite{Parisi:1980ys,Parisi:1984cs,Cristoforetti:2012su,Fujii:2013sra,Mori:2017pne,Mori:2017nwj} for recent progresses in methods to tackle the sign problem as an example.
Therefore, several expectations at moderate $\mu_\mathrm{R}$ have been obtained by using QCD effective models; see Refs.\,\cite{Fukushima:2010bq,Buballa:2016fjh} as an example.

In the early stage of the study for the QCD phase diagram, the first-order transition was expected to be appeared at moderate $\mu_\mathrm{R}$ even with sufficiently low $T$.
However, the duality between the hadron phase and the color superconducting phase, which include not only the color-flavor locking (CFL) phase but also the two-flavor color-superconducting (2SC) phase, has been proposed~\cite{Schafer:1998ef,Alford:1999pa,Fujimoto:2019sxg}; there are one to one correspondence of elementary excitation modes between phases.
It is the so called the quark-hadron continuity. 
In addition, there are several studies that predict the crossover
between the hadron phase and the deconfined quark matter by using the QCD effective model~\cite{Kitazawa:2003qmg} and also the Ginzburg-Landau analysis~\cite{Hatsuda:2006ps}.
Of course, the duality is not confirmed yet; see Refs.\,\cite{Alford:2018mqj,*Chatterjee:2018nxe,*Hirono:2018fjr} for recent progresses.

To understand the QCD phase diagram at moderate $\mu_\mathrm{R}$, the quakyonic phase~\cite{McLerran:2007qj} may play a crucial role.
The quarkyonic phase is first proposed in the large $N_\mathrm{c}$ QCD by using the $N_\mathrm{c}$ counting where $N_\mathrm{c}$ is the number of color; see Refs.\,\cite{tHooft:1973alw,*tHooft:1974pnl,Witten:1979kh}.
In the large $N_\mathrm{c}$, quarks can be treated as the probe if $\mu_\mathrm{R}$ is not reached to ${\cal O}(N_\mathrm{c})$.
Then the pressure is ${\cal O} (N_\mathrm{c}^0)$ in the confined phase because the physical degree of freedoms are the gluballs and baryons, but it does ${\cal O} (N_\mathrm{c}^2)$ in the deconfined phase because physical degree of freedoms are gluons and quarks.
Interestingly, the quark number density start to have nonzero value when $\mu_\mathrm{R}$ reaches to $M_\mathrm{B}/N_\mathrm{c}$ where $M_\mathrm{B}$ is the lowest baryon mass.
In this case, the pressure is ${\cal O}(N_\mathrm{c}^1)$, and it can be interpreted as follows:
The thermodynamic quantities are dominated by quarks inside the Fermi sea, but the physical excitation modes on the Fermi surface are corresponding to baryonic because the confined nature is not changed; the quarks are probe and thus they can not modify the gauge field configuration.
Recently, the quarkyonic phase has also been investigated by using the top-down approach based on the AdS/CFT correspondence~\cite{Kovensky:2020xif}.
Unfortunately, the quakyonic phase is not clear when we consider small $N_\mathrm{c}$ such as $N_\mathrm{c}=3$, but some discussions have been done with QCD effective models; for example, see Ref.\,\cite{McLerran:2008ua}. 
There is the discussion that the quarkyonic phase still important in the $N_\mathrm{c}=3$ system such as the new confinement-deconfinement transition scenario~\cite{Fukushima:2020cmk}.
In addition, the quarkyonic phase may affect the neutron star properties such as the mass-radius relation~\cite{McLerran:2018hbz}.

To understand the physical degree of freedom, the canonical ensembles may provide important information since they are directly related with each quark number.
Therefore, we employ the canonical ensemble method~\cite{Alexandru:2005ix,Kratochvila:2006jx,deForcrand:2006ec,Fukuda:2015mva,Oka:2017kny} to discuss the QCD phase structure at moderate $\mu_\mathrm{R}$ with low $T$ in this paper.
The canonical sectors are constructed by using the imaginary chemical potential ($\mu_\mathrm{I}$) and then several knowledge of QCD at finite $\mu_\mathrm{I}$ play a crucial role; see Refs.~\cite{Roberge:1986mm,D'Elia:2002gd,deForcrand:2002ci,deForcrand:2003hx,D'Elia:2004at,Kashiwa:2017swa,Kashiwa:2019ihm} as an example.
In this paper, we first consider the quakyonic phase.
The color superconducting phase is, of course, interesting phase on the QCD phase diagram, but the gauge symmetry issue in the canonical ensemble method is a tricky issue.
Thus, it is difficult to investigate the phase in the present approach, but we show some qualitative discussions for the color superconducting phase with some ansatz in this paper.
In addition, the chiral symmetry restoration with increasing $\mu_\mathrm{R}$ is discussed from the canonical sectors.

This paper is organized as follows.
In the next section\,\ref{Sec:CEM}, we explain the formulation of the canonical ensemble method.
Section\,\ref{Sec:large_nc} explains the discussions about the canonical sectors in the large $N_\mathrm{c}$ and Sec.\,\ref{Sec:example} shows the simple estimation of the canonical sectors by using the Polyakov-loop extended Nambu--Jona-Lasinio model as an example.
The chiral properties and the color superconducting are discussed in Sec.\,\ref{Sec:chiral} and \ref{Sec:diquark}, respectively.
Section\,\ref{Eq:summary} is devoted to the summary.

\section{Canonical ensemble method}
\label{Sec:CEM}

Throughout whole discussions in this paper, we consider large but finite size continues system because the thermodynamic limit requires careful treatment for the infinite sums.
Starting from the grand-canonical partition function (${\cal Z}_\mathrm{GC}$) at finite $T$ and $\theta:=\mu_\mathrm{I}/T$, we can construct the canonical partition function (${\cal Z}_\mathrm{C}$) with fixed quark number ($Q$) as
\begin{align}
    {\cal Z}_\mathrm{C} (Q)
    &= \sum_n \bra{n} e^{-\beta {\cal H}} \delta (\hat{n}-Q) \ket{n} 
\nonumber\\
    &= \frac{1}{2\pi} \int_{-\pi}^\pi
       e^{i Q \theta} {\cal Z}_\mathrm{GC} (\theta) \, d\theta
       \nonumber\\
    &= \frac{1 + z^Q + \cdots + z^{Q(N_\mathrm{c}-1)}}{2\pi}
       \int_{-\pi/N_\mathrm{c}}^{\pi/{N_\mathrm{c}}}
       e^{i Q \theta} {\cal Z}_\mathrm{GC} (\theta) \, d\theta
       \nonumber\\
    &=
    \begin{dcases}
    \frac{N_\mathrm{c}}{2\pi}  
    \int_{-\pi/N_\mathrm{c}}^{\pi/{N_\mathrm{c}}}
       e^{i Q \theta} {\cal Z}_\mathrm{GC} (\theta) \, d\theta
       & (Q=3 k)\\
     ~0 & (Q \neq 3k)
    \end{dcases}
    ,
\label{Eq:canonical}
\end{align}
where ${\cal H}$ menas the Hamiltonian, $k \in \mathbb{Z}$, $z = \exp(2\pi i/N_\mathrm{c})$ is the ${\mathbb Z}_{N_\mathrm{c}}$ factor and $\hat{n}$ is the quark number operator.
In this paper, we do not explicitly show $T$ for the argument of the partition function and also other quantities because we are interested in $\mu_\mathrm{R}$ effects.
With the fugacity expansion, we have
\begin{align}
    {\cal Z}_\mathrm{GC}(\mu_\mathrm{R})
    &= \sum_n \bra{n} e^{ - (\beta H  -\mu_\mathrm{R} \hat{n} ) }
       \ket{n}
\nonumber\\
    &= {\cal Z}_\mathrm{C}(0)
     + e^{\frac{\mu_\mathrm{R}}{T}}
       {\cal Z}_\mathrm{C}(N_\mathrm{c})
     + \cdots
\nonumber\\
    &= \sum_{n_\mathrm{B}=-\infty}^\infty
       e^{n_\mathrm{B} \frac{\mu_\mathrm{B}}{T}}
       {\cal Z}_\mathrm{C}(N_\mathrm{c} n_\mathrm{B}),
\label{Eq:fugacity}
\end{align}
where $n_\mathrm{B}=n / N_\mathrm{c}$ is the baryon number and $\mu_\mathrm{B} = N_\mathrm{c} \mu_\mathrm{R}$ does the baryon chemical potential; we here use the fact that $N_\mathrm{c}$ multiples of $n$ only contribute ${\cal Z}_\mathrm{C}$ because of the Roberge-Weiss (RW) periodicity; see Eq.\,(\ref{Eq:canonical}) and Ref.\,\cite{Roberge:1986mm} as an example.
By using the above relations and the lattice QCD data at finite $\mu_\mathrm{I}$, we can investigate the QCD phase structure at finite $\mu_\mathrm{R}$ with certain $T$ where the numerical error induced from the Fourier transformation can be controlled~\cite{Alexandru:2005ix,Kratochvila:2006jx,deForcrand:2006ec,Fukuda:2015mva,Oka:2017kny}.
It is noted that above expression is valid even for low $T$ and high $\mu_\mathrm{R}$; numerical confirmations of it in QCD effective models can be seen in Refs.\,\cite{Wakayama:2019hgz,Wakayama:2020dzz}.
Therefore, ${\cal Z}_\mathrm{GC} (\mu_\mathrm{R})$ is constructed from ${\cal Z}_\mathrm{GC} (\theta)$, and  vice versa.
Finally, we have the inverse relation as
\begin{align}
    {\cal Z}_\mathrm{GC}(\theta)
    &= \sum_{n_\mathrm{B}=-\infty}^\infty e^{-i  n_\mathrm{B} N_\mathrm{c} \theta} {\cal Z}_\mathrm{C}(N_\mathrm{c} n_\mathrm{B})
\nonumber\\
    &= {\cal Z}_\mathrm{C}(0) + 2\sum_{n_\mathrm{B}=1}^\infty {\cal Z}_\mathrm{C}(N_\mathrm{c} n_\mathrm{B}) \cos(N_\mathrm{c} n_\mathrm{B} \theta).
\end{align}
This relation means that ${\cal Z}_\mathrm{GC}(\theta)$ is decomposed by the canonical sectors (oscillating modes) and thus the canonical sectors survey elementary excitation modes via the oscillating behaviors. 
This fact has been used to investigate the confinement-deconfinement nature at finite $T$ with $\mu_\mathrm{R}=0$~\cite{Kashiwa:2017swa}.

\section{Large $N_\mathrm{c}$ estimation}
\label{Sec:large_nc}

\begin{figure}[t]
 \centering
 \includegraphics[width=0.5\textwidth]{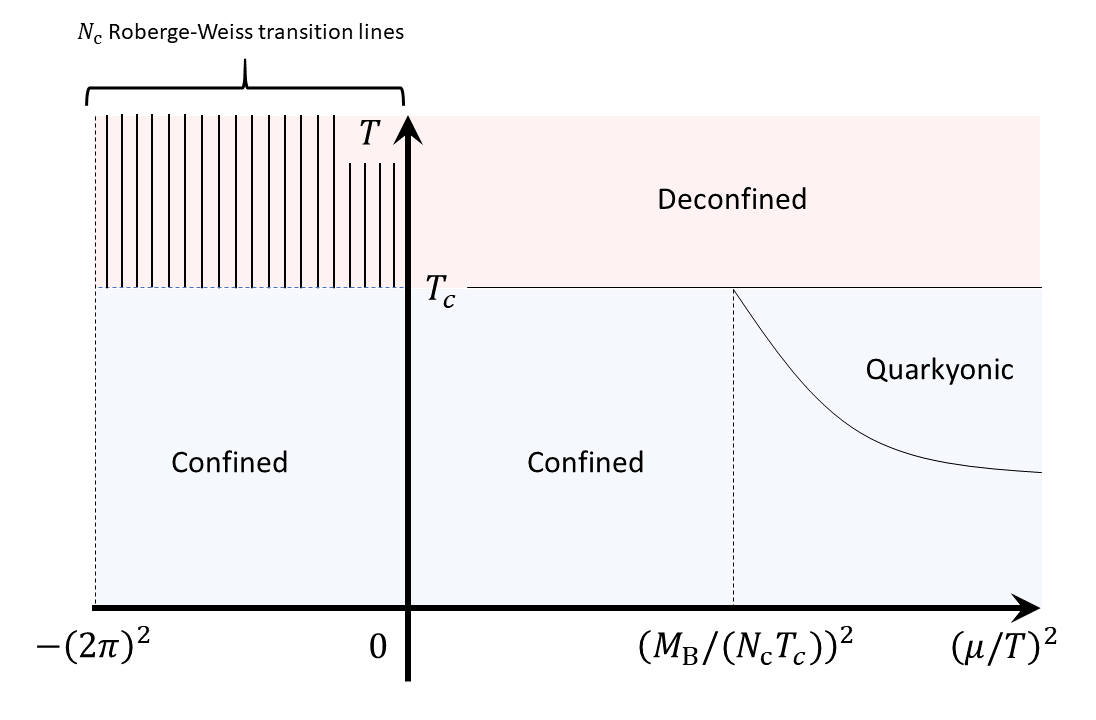}\\
 \includegraphics[width=0.5\textwidth]{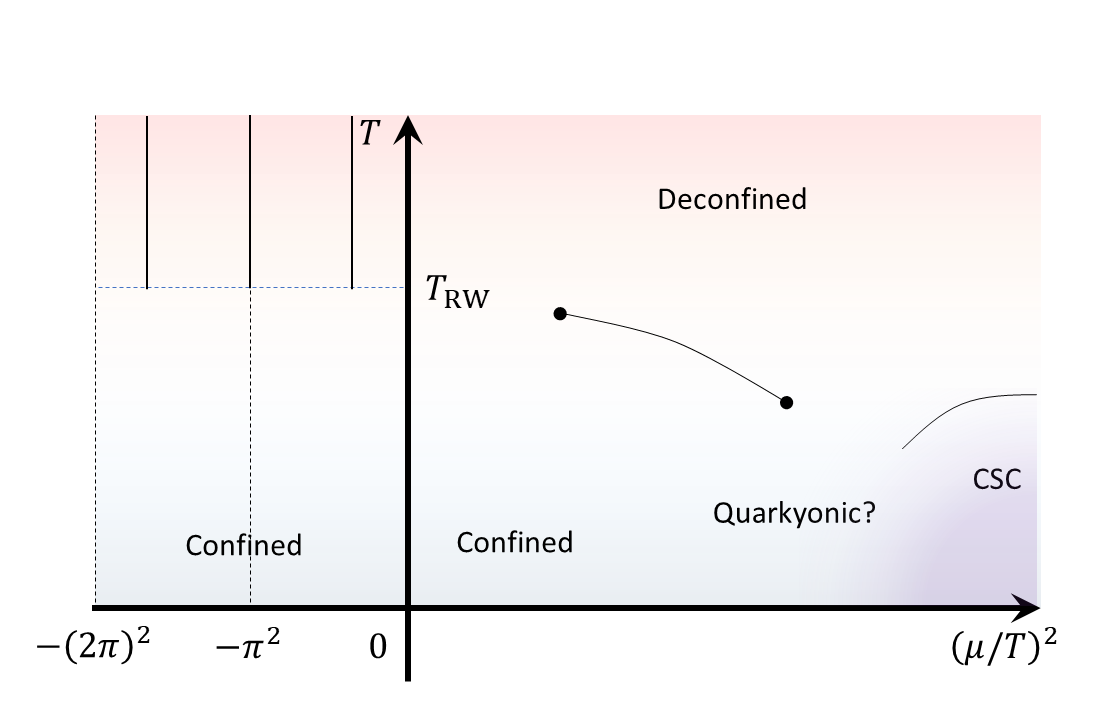}
 \caption{
 The schematic figures of the QCD phase diagram in the $(\mu/T)^2$-$T$ plane with sufficiently large $N_\mathrm{c}$ (top) and that with $N_\mathrm{c}=3$ (bottom).
 Solid lines represent the phase transition lines and closed circles mean the critical endpoint.
 The legend CSC means the color superconducting phase such as the CFL and 2SC and $T_\mathrm{RW}$ denots the Roberge-Weiss endpoint temperature which is almost corresponding to the deconfinement critical temperature at $\mu=0$ with large $N_\mathrm{c}$.
 In the bottom panel, the confined phase is not clear meaning above the liquid-gas transition which is not explicitly shown in the figure.
 Also, we assume that there is no chiral phase transition at low $T$ in the figure; it is, of course, not confirmed yet.
}
\label{Fig:Phase_Diagram}
\end{figure}

Via the $N_\mathrm{c}$ counting of meson, baryon, gluball, gluon and quark contributions, the quakyonic matter has been proposed in the large $N_\mathrm{c}$ QCD; see Fig.\,\ref{Fig:Phase_Diagram}.
Since quark loops are $1/N_\mathrm{c}$ suppressed, the oscillation of quantities at finite $\theta$ below the critical temperature $T_\mathrm{c}$ may be approximated by simple $\cos (N_\mathrm{c} \theta)$ function.
Therefore, we can assume
\begin{align}
    {\cal Z}_\mathrm{GC}(\theta) &= a + b_1 \cos (N_\mathrm{c} \theta),
\label{Eq:assumption}
\end{align}
where $a$ and $b_1$ should depend on $T$ and the spatial volume ($V$). 
For example,  the similar functional form of Eq.\,(\ref{Eq:assumption}) is obtained in the strong coupling calculation; see Refs.\,\cite{Nishida:2003fb,Kawamoto:2005mq}.

With decreasing $T$ even at small $N_\mathrm{c}$, the oscillation in terms of $\theta$ becomes weak and thus $b_1 \to 0$ with $T \to 0$ because $\theta$ can be transformed as the temporal boundary condition of quarks.
Higher order terms are expected to be suppressed and thus we neglected here.
The first term in Eq.\,(\ref{Eq:assumption}), $a$, represents the gluball contributions because it does not have the $\mu$-dependence.
The second term should mainly contain the baryonic contributions.

In this work, we consider large but finite $N_\mathrm{c}$, $1 \ll N_\mathrm{c} < \infty$, from following reasons:
The phase rotation of the Polyakov loop in terms of $\theta$ does not happen in the large $N_\mathrm{c}$ limit because $\mathbb{Z}_{N_\mathrm{c}}$ broken quark contributions, which induces the RW periodicity, can not be modified $\mathbb{Z}_{N_\mathrm{c}}$ symmetric gluon contributions.
This indicates that the period in terms of $\theta$ is $2\pi$ at least in the high $T$ region as mentioned in Refs.\,\cite{Doi:2017dyc,Doi:2017jad,Ghoroku:2020fkv}, which is shape contrast with the finite $N_\mathrm{c}$ case, when quarks become the probe.
At low $T$, the system is perfectly dominated by gluballs and baryons in the large $N_\mathrm{c}$ limit and then the periodicity issue may be relaxed; there is the RW periodicity.
However, the period of the RW periodicity at low $T$ becomes $0$ in the large $N_\mathrm{c}$ limit and then it is highly nontrivial that the Fourier transformation is well defined or not when we consider the order of operation about $N_\mathrm{c} \to \infty$ and the integration of Fourier transformation; see Sec.\,\ref{Sec:CEM}.
Therefore, we need the infinitesimally small but nonzero  back-reaction of quarks to the gluon contributions in this work.

Via the change of variables in the partition function, $\theta$ can be absorbed into the temporal boundary condition of quarks and thus it is irrelevant at low $T$. 
This indicates the important consequence:
If the system is confined at $\mu=0$ with fixed $T$, the imaginary chemical potential region should be the confined region.
Then, the canonical sectors constructed by using the imaginary chemical potential should be independent of $\mu_\mathrm{R}$ and thus all canonical sectors only have the confined information it should be.
This fact is valid even if $N_\mathrm{c}$ is small if the confined nature is strong enough such as the $T \sim 0$ situation.
Below, we consider sufficiently low $T$ because both the fugacity and ${\cal Z}_\mathrm{C}(k)$ with $k=1,2,\cdots$ depend on $T$ and thus discussions about the $T$-dependence is very difficult unlike the $\mu_\mathrm{R}$-dependence.

The canonical sectors with Eq.\,(\ref{Eq:assumption}) become
\begin{align}
    Z_\mathrm{C}(0) = a,~~~~
    Z_\mathrm{C}(N_\mathrm{c}) = b_1,
\end{align}
because
\begin{align}
    \int_{-\pi/N_\mathrm{c}}^{\pi/N_\mathrm{c}} d\theta \, e^{i n \theta} \cos(m\theta) &=
    \begin{dcases}
    ~0 & (n \neq m)\\
    ~\mathrm{nonzero} & (n=m)
    \end{dcases}
.
\label{Eq:integral}
\end{align}
These results are, of course, consequences from the properties of Fourier transformation; we put them here to show the consistency.
Then,we have
\begin{align}
    {\cal Z}_\mathrm{GC}(\mu_\mathrm{R})
    &= a + b_1 e^{ N_\mathrm{c} \frac{\mu_\mathrm{R}}{T}}
         + b_1 e^{-N_\mathrm{c} \frac{\mu_\mathrm{R}}{T}}.
\end{align}
Below, we consider large $N_\mathrm{c}$ and nonzero positive $\mu_\mathrm{R}$ and thus we neglect the third term. 
Since the second term depends on $\mu_\mathrm{R}$ and thus there is the region where the first and the second term are balanced at ${\tilde \mu}_\mathrm{R}$;
\begin{align}
    e^{ N_\mathrm{c} \frac{{\tilde \mu}_\mathrm{R}}{T}}
    &= \Bigl| \frac{a}{b_1} \Bigr|.
\label{Eq:balance}
\end{align}
Since the second term is the $N_\mathrm{c}$ quarks (baryon) contribution and thus this energy scale may be related with the quarkyonic phase transition.
It is well known that the quakyonic phase transition is happen around $\mu_\mathrm{R} = M_\mathrm{B}/ N_\mathrm{c}$.
When we match this value with Eq.\,(\ref{Eq:balance}), we have
\begin{align}
    e^{ N_\mathrm{c} \frac{{\tilde \mu}_\mathrm{R}}{T}}
    &= e^{ N_\mathrm{c} \frac{M_\mathrm{B}}{N_\mathrm{c}T}}
     =  \Bigl| \frac{a}{b_1} \Bigr|.
\end{align}
Therefore, we obtain
\begin{align}
    |b_1| 
    &= |a| \,
       e^{ -\frac{M_\mathrm{B}}{T}}.
\label{Eq:b_1}
\end{align} 
This result seems to be natural because of following two reasons.
First, $b_1$ comes from the baryon contribution in the case and thus appearance of $M_\mathrm{B}$ is natural.
Second, the oscillation of ${\cal Z}_\mathrm{GC}$ in terms of $\theta$ should be vanished with $T \to 0$, and this fact is realized in Eq.\,(\ref{Eq:b_1}) as $b_1 \to 0$ with $T \to 0$.
In addition, we can expect this kind of the functional form is qualitatively obtained in the QCD effective model with suitable simplifications; see the next section~\ref{Sec:example}.

The pressure ($p$) is obtained from the thermodynamic relation as
\begin{align}
    p(\mu_\mathrm{R}) &= \frac{1}{\beta V} \ln {\cal Z}_\mathrm{GC}(\mu_\mathrm{R}),
    \label{Eq:pressure}
\end{align}
where $\beta$ is the inverse temperature, $\beta = 1/T$.
The $b_1$ term induces the $N_\mathrm{c}^1$-order contributions because $M_\mathrm{B} \propto N_\mathrm{c}$ until $\mu_\mathrm{R}$ reaches $N_\mathrm{c}^1$-order.
The ${\cal Z}_\mathrm{C}(0)$ term does not have the $\mu_\mathrm{R}$-dependence but $e^{N_\mathrm{c} \mu_\mathrm{R}/T} {\cal Z}_\mathrm{C}(N_\mathrm{c})$ have the dependence and thus the $b_1$ term leads the opening contribution to nonzero quark number density ($n_q$);
\begin{align}
    n_q(\mu_\mathrm{R}) &= \frac{1}{\beta V} \frac{\partial \ln {\cal Z}_\mathrm{GC}(\mu_\mathrm{R}) }{\partial \mu_\mathrm{R}}.
\end{align}
Therefore, $n_q \sim 0$ remains until $\mu_\mathrm{R} \sim M_\mathrm{B}/N_\mathrm{c}$ and it is turned into $N_\mathrm{c}^1$-order above the value with large $N_\mathrm{c}$ and low $T$.
This behavior is also consistent with the quarkyonic picture in large $N_\mathrm{c}$ QCD about the quark number density.

Finally, in this section, we discuss following two scenarios.
Via the present treatment used in this study, at least, we discuss precursory phenomenon for the phase transitions.
The first one is (A): the change of infinite tower of the canonical sectors with increasing $\mu_\mathrm{R}$ and the other is (B): the nonexistence of the change. 
\\
\begin{description}
   \item[Scenario (A)] With the large $N_\mathrm{c}$, ${\cal Z}_\mathrm{C} (0)$ (${\cal Z}_\mathrm{C} (N_\mathrm{c})$) dominate $\mu_\mathrm{R} < {\tilde \mu}_\mathrm{R}$ ($\mu_\mathrm{R} > {\tilde \mu}_\mathrm{R}$).
Since $a$ should be the gluball contribution and thus it leads $N_\mathrm{c}^0$ contributions to the pressure.
Interestingly, there is the balanced region ${\tilde \mu}_\mathrm{R} - \epsilon < \mu_\mathrm{c}< {\tilde \mu}_\mathrm{R} + \epsilon$ with $\epsilon \sim \Lambda_\mathrm{QCD}/N_\mathrm{c}$ where we use $\Lambda_\mathrm{QCD}$ in the order counting.
If each coefficient has, at least, the $1/n$ suppression factor, the dominant canonical sectors are changed as
\begin{align}
    {\cal Z}_\mathrm{C}(0) \to {\cal Z}_\mathrm{C}(N_\mathrm{c}) \to {\cal Z}_\mathrm{C}(2 N_\mathrm{c}) \to \cdots,
\end{align}
where the infinite tower of the canonical sectors are spanned with increasing $\mu_\mathrm{R}$ because of the exponential suppression factors for each canonical sector.
Since quark loops are suppressed in the $N_\mathrm{c} \to \infty$ limit and thus we can naturally expect such $1/n$ suppression factor; we may expect stronger suppression factor from the lattice QCD data at finite $T$~\cite{Nagata:2012pc}. If we have stronger suppression factor, discussions in this paper are unchanged.
In the large $N_\mathrm{c}$, the balanced region is very tiny, but it is enlarged when $N_\mathrm{c}$ becomes smaller.
This indicates that there is no clear phase transition in the realistic $N_\mathrm{c}$ at moderate $\mu_\mathrm{R}$ with low $T$.
This expectation is consistent with recent discussion for the moderate $\mu_\mathrm{R}$ region; there is no phase transition between the nuclear and quark matters.
In contrast, in the case of realistic $N_\mathrm{c}$ case, it is difficult to obtain clear consideration, but we can expect the balanced region is enlarged from the above estimation of the region and thus we do not have clear quakyonic phase transition energy scale; see the bottom panel of Fig.\,\ref{Fig:Phase_Diagram} as a schematic figure of the QCD phase diagram.
This scenario is most likely scenario in QCD.
   \item[Scenario (B)] It should be noted that all higher-order canonical sectors start to contribute the system at ${\tilde{\mu}}_\mathrm{R}$ if each higher-order canonical sector does not have the $1/n$-type suppression factor.
   Then, we can not have the infinite tower picture unlike the scenario (A). In this case, we may have the sharp energy scale of the quarkyonic phase transition even in the realistic $N_\mathrm{c}=3$ system form the comparison between ${\cal Z}_\mathrm{GC}(0)$ and ${\cal Z}_\mathrm{GC}(N_\mathrm{c})$.
   However, we can find the suppression factor from the simple estimation by using the QCD effective model as in the next section \ref{Sec:example} and thus this scenario is not likely scenario in QCD.
   It should be noted that ${\cal Z}_\mathrm{GC}$ is not simply converged in this scenario with increasing $n$ and thus it seems to be rejected from the mathematical sense.
\end{description}

\section{Simple model estimation}
\label{Sec:example}

In this section, we show canonical sectors estimated by employing the Polyakov-loop extended Nambu--Jona-Lasinio (PNJL) model~\cite{Fukushima:2003fw,Fukushima:2017csk} as an example.
The PNJL model is the extended model of the NJL model to include the Polyakov-loop dynamics by considering the mean-field of the temporal component of the gluon field ($A_4$) with the homogeneous ansatz.
Of course, it does not contain the exact confinement mechanism, but it can mimic not only the approximated confinement nature but also several important properties such as the RW periodicity and its transition which are closely related with physical degree of freedom of the system~\cite{Kashiwa:2015tna,Kashiwa:2016vrl}.
Therefore, we here employ the PNJL model to estimate canonical sectors and show the estimation is matched with the result obtained in the previous section~\ref{Sec:large_nc}.

In the following estimation, we assume that the constituent quark mass ($M$) is order of the QCD energy scale ($\Lambda_\mathrm{QCD}$) in the finite $\theta$ region with low $T$ when we estimate the Fourier transformation.
In the PNJL model, we usually employ the Polyakov-gauge fixing, $\partial A_4 =0$, and then $A_4$ is diagonalized by using the gauge degree of freedom remained in the spatial component.
The Lagrangian density of the two-flavor PNJL model is given by
\begin{align}
{\cal L}
&= {\bar q} (i \gamma^\mu D_\mu - m_0 ) q + G[({\bar q}q)^2 + ({\bar q} i \gamma_5 \vec{\tau} q)^2]
 - {\cal U},
\end{align}
where $m_0$ denotes the current quark mass, $D_\mu$ is the covariant derivative $D^\mu = \partial^\mu + i \delta^{4}_\mu A^\mu$, $G$ is the coupling constant and ${\cal U}$ does the gluonic contribution.
In this paper, ${\cal U}$ is not important and thus we do not show the explicit functional form; for example, see Refs.\,\cite{Fukushima:2003fw,Ratti:2005jh,Roessner:2006xn} With the mean-field approximation, the effective potential is given by
\begin{align}
    {\cal V} 
    &= -2 N_\mathrm{f}
       \int \frac{dp \, p^2}{4\pi^2}
       \Bigl[ N_\mathrm{c}E - T \sum_{\eta=\mp 1}\ln \det (1+e^{-\beta (E + \eta {\tilde \mu})}) \Bigr]
    \nonumber\\
    & \hspace{0.4cm} + G \sigma^2 + {\cal U},
    \nonumber\\
    &= -N_\mathrm{f}
       \int \frac{dp \, p^2}{2\pi^2}
       \Bigl[ N_\mathrm{c}E + T \ln (f^- f^+) \Bigr]
        + G \sigma^2 + {\cal U},
\label{Eq:ep1}
\end{align}
where $N_\mathrm{f}=2$, ${\tilde \mu} = \mu + i A_4$ and
\begin{align}
    f^- &= 1 + (\Phi + {\bar \Phi}e^{-\beta E^- }) \, e^{-\beta E^- } + e^{- N_\mathrm{c} \beta E^- },
\nonumber\\
    f^+ &= 1 + ({\bar \Phi} + \Phi e^{-\beta E^+ }) \, e^{-\beta E^+ } + e^{- N_\mathrm{c} \beta E^+},
\label{Eq:ep_PNJL}
\end{align}
here $E^\mp=E\mp\mu$, $\sigma = \langle {\bar q} q \rangle$ and $E=\sqrt{p^2+M^2}$ with $M=m-2 G \sigma$.
The definition of the Polyakov loop ($\Phi$) and its conjugate (${\bar \Phi}$) in the model are
\begin{align}
    \Phi &= \frac{1}{N_\mathrm{c}} \mathrm{tr_c} \, e^{i\beta \langle A_4 \rangle},~~
    {\bar \Phi} = \frac{1}{N_\mathrm{c}} \mathrm{tr_c} \, e^{-i\beta \langle A_4 \rangle},
\end{align}
where $\mathrm{tr_c}$ is the trace acts on the color space.
The above effective potential (\ref{Eq:ep_PNJL}) is corresponding to the leading-order result of the $1/N_\mathrm{c}$ expansion and then the higher order contributions which are corresponding to the meson loop contributions are neglected; see Ref.\,\cite{Kashiwa:2003rj} for the result in the NJL model as an example.

Since we here consider the low $T$ region and thus the Polyakov-loop dynamics is usually decoupled from the system; we here consider the $T$ region where the Polyakov loop and some other higher-order loops are sufficiently weaker than $\exp(-N_\mathrm{c} \beta (E \mp \mu))$.
Therefore, the effective potential (\ref{Eq:ep1}) may be simplified as
\begin{align}
    {\cal V} 
    &= - N_\mathrm{f}
       \int \frac{dp \, p^2}{2 \pi^2}
       \Bigl[ N_\mathrm{c} E
            + T \ln ({\tilde f}^- {\tilde f}^+)
        \Bigr]
    \nonumber\\
    &\hspace{0.4cm} + G \sigma^2 + {\cal U},
    \label{eq:approx_eq_0}
\end{align}
where
\begin{align}
    {\tilde f}^\mp &= 1 + e^{- N_\mathrm{c} \beta (E \mp \mu) }.
\end{align}
It should be noted that we here consider finite imaginary chemical potential and thus $\mu= i\mu_\mathrm{I}$.
When we consider $N_\mathrm{c}>3$, there should be contributions of higher order loops~\cite{Meisinger:2001cq} and thus $f^\mp$ must be modified comparing with those in $N_\mathrm{c}=3$, but we can expect the appearance of the $e^{- N_\mathrm{c} \beta (E+\mu) }$ term in the logarithmic part because of properties of $SU(N_\mathrm{c})$ Lie group;
it is because $\ln \det$ term in Eq.\,(\ref{Eq:ep1}) must have $\exp[- N_\mathrm{c} \beta (E \mp \mu)]$.
This term mimics the $N_\mathrm{c}$-quark state and the suppression of other $m$-quark states with $m=1,\cdots, N_\mathrm{c}-1$ at low $T$ does the QCD confinement nature; see Ref.\,~\cite{Sasaki:2006ww} for discussions in the case of $N_\mathrm{c}=3$. 
In addition, $T$ is set to be sufficiently small and thus we can approximate Eq.\,(\ref{Eq:ep1}) as
\begin{align}
    {\cal V} 
    &= -N_\mathrm{f} T
       \int \frac{dp \, p^2}{2 \pi^2}
       \Bigl[ e^{- N_\mathrm{c} \beta (E-\mu) } + e^{- N_\mathrm{c} \beta (E+\mu) }
        \Bigr]
    \nonumber\\
    & \hspace{0.4cm} + a,
    \label{Eq:approx_eq}
\end{align}
where $\sigma$ depends on $\theta$, but the dominant part can be expected as $\theta$-independent and thus we approximate the second and third terms in Eq.\,(\ref{eq:approx_eq_0}) into the $\theta$-independent $a$.
In the case of the large $N_\mathrm{c}$, $N_\mathrm{c} M$ dominates the integration (static limit) as
\begin{align}
    p^2 e^{-\beta N_\mathrm{c} \sqrt{p^2 + M^2}}
    \to \epsilon^2 e^{-\beta N_\mathrm{c} \sqrt{\epsilon^2 + M^2}},
\end{align}
because the exponential suppression is much stronger than $p^2$ in the large $N_\mathrm{c}$, and
\begin{align}
    N_\mathrm{c} \sqrt{{\tilde \epsilon}^2 + M^2} \gg N_\mathrm{c} M,
\end{align}
where $\epsilon$ and ${\tilde \epsilon}$ are the infinitesimal value proportional to $1/N_\mathrm{c}$ and the infinitesimal value which manifests ${\tilde \epsilon} \gg 1/N_\mathrm{c}$.
Therefore, we consider the momentum integration up to the region where we do not reach ${\bar \epsilon}$ scale because of the strong exponential suppression effect.
Thus, we have
\begin{align}
    {\cal V} \sim a + b \, e^{-\beta N_\mathrm{c} M} \cos(N_\mathrm{c} \theta)+ \cdots,
\end{align}
where $a$ is the $\theta$-independent and $b$ does the $\theta$-dependent coefficients.
Here, we use $M \sqrt{\epsilon^2/M^2 + 1} \sim M$.
If the saddle-point approximation is good, we can estimate the grand-canonical partition function as
\begin{align}
    {\cal Z}  \propto e^{- \beta V {\cal V}}.
\end{align}
Therefore, we finally obtain
\begin{align}
    {\cal Z}
    &\sim e^{-\beta V a}
          \Bigl[ 1 -\beta V b \, e^{-N_\mathrm{c} M /T} \cos (N_\mathrm{c} \theta) +\cdots \Bigr]
    \nonumber\\
    & = {\tilde a} + {\tilde b} \, e^{-N_\mathrm{c} M /T} \cos(N_\mathrm{c} \theta) + \cdots,
    \label{Eq:pf_qem}
\end{align}
with sufficiently low $T$ and fixed $V$.
The second term is corresponding to $b_1$ in the previous section.
Interestingly, although we use too much simplification, we can find the $\exp(-N_\mathrm{c} M / T) \sim \exp(-M_\mathrm{B}/T)$ factor in the present estimation.
This form is exactly the same one that we obtained in the large $N_\mathrm{c}$ estimation via the matching with quarkyonic phase transition energy scale in Sec.\,\ref{Sec:large_nc};
the difference of coefficients between the present and previous estimation can be adjusted by $\mu_\mathrm{c} \to \mu_\mathrm{c} + \alpha$ and $\alpha$ may be order ${\cal O} ( \frac{\ln N_\mathrm{c}}{N_\mathrm{c}} )$. 

In principle, higher-order contributions can exist in Eq.\,(\ref{Eq:pf_qem}) and then we have $\exp(- m N_\mathrm{c} M/T)$ terms with $m=1,2,\cdots$; those are not important in the region where the dominant oscillating mode is $\cos(N_\mathrm{c} \theta)$ when $\mu_\mathrm{R}$ is not large enough.
It should be noted that we expand the logarithmic functions by $\exp(- N_\mathrm{c} M/T)$ in Eq.\,(\ref{Eq:approx_eq}) and thus higher-order expansion terms should have, at least, the $1/n$ suppression factor; existence of this factor is assumed in the previous section, but we can find it in the present estimation.

\section{Chiral properties}
\label{Sec:chiral}

In this section, we discuss the chiral properties from the canonical sectors.
In the case of the chiral condensate, it can have finite value at finite $\theta$ at low $T$.
Then, the chiral condensate at finite $\mu_\mathrm{R}$ can be expressed as
\begin{align}
    \langle \sigma \rangle (\mu_\mathrm{R})
    &= \sum_{n=-\infty}^{\infty}
    e^{n\mu_\mathrm{R}/T}
       \frac{{\cal Z}_\mathrm{C}(N_\mathrm{c} n)}{{\cal Z}_\mathrm{GC}(\mu_\mathrm{R})}
       \langle \sigma \rangle_\mathrm{C} (N_\mathrm{c} n).
\end{align}
At low $\mu_\mathrm{R}$, higher-order canonical sectors than the $n=0$ sector are irrelevant and then the contribution of $\langle \sigma \rangle_\mathrm{C} (0)$ dominate $\langle \sigma \rangle (\mu_\mathrm{R})$.

It is known that the absolute value of the chiral condensate at finite $\theta$ becomes larger than that at $\mu=0$ from the lattice QCD simulation~\cite{D'Elia:2002gd} and the QCD effective model calculations~\cite{Sakai:2008py}.
This indicates that the oscillating behavior at sufficiently low $T$ is expected as
\begin{align}
    \langle \sigma \rangle (\theta) &= a_\sigma - b_\sigma \cos (N_\mathrm{c} \theta),
\end{align}
where $a_\sigma$ and $b_\sigma$ are positive coefficients and $\sigma$ means the absolute value of the chiral condensate.
In the $T \to 0$ limit, $a_\sigma$ dominates the system and then we can expect $a_\sigma$ is the $\Lambda_\mathrm{QCD}$-order; the first term $\langle \sigma \rangle_\mathrm{C} (0)$ can be assumed as the $\Lambda_\mathrm{QCD}$-order in the setup.
When $T$ increases, higher-order oscillating modes start to appear in the equation. 
Therefore, the canonical sectors with $n=N_\mathrm{c}$ should have the opposite sign of the canonical sector with $n=0$ after the Fourier transformation because of Eq.\,(\ref{Eq:integral}) as
\begin{align}
& \mathrm{sgn} \Bigl[ \int_{-\pi/N_\mathrm{c}}^{\pi/N_\mathrm{c}}
    e^{i n \theta} \{a_\sigma - b_\sigma \cos (N_\mathrm{c} \theta) \} d\theta \Bigr]_{n=0} = +1,
\nonumber\\
&   \mathrm{sgn} \Bigl[ \int_{-\pi/N_\mathrm{c}}^{\pi/N_\mathrm{c}}
    e^{i n \theta} \{a_\sigma - b_\sigma \cos(N_\mathrm{c} \theta)\} d\theta
    \Bigr]_{n=N_\mathrm{c}} = -1,
    \nonumber
\end{align}
where $\mathrm{sgn}$ is the sign function.
Because of the sign difference, the $n=0$ canonical sector is weaken by the $n = N_\mathrm{c}$ canonical sector.
If $\mu_\mathrm{R}$ increases more and more, higher-order canonical sectors join the cancellation of the $n=0$ canonical sector and then the sign for each canonical sectors become important.
To discuss more details, we need actual numerical data and thus it is left unclear problem which will be clarify in our future work.
Present discussion is the picture of the chiral symmetry restoration with increasing $\mu_\mathrm{R}$ at sufficiently low $T$ from the viewpoint of the canonical sectors.

If the chiral phase transition which can be characterized by the corresponding local order-parameter happens at $\mu_\mathrm{R}=\mu_\sigma$, ${\cal Z}_\mathrm{GC}$ becomes zero very close to $\mu_\sigma$ on the complex $\mu$ plane; for example, this fact plays a crucial role in the Lee-Yang zero analysis~\cite{yang1952statistical,*lee1952statistical}. 
With sufficiently low $T$, we can estimate the transition point from
\begin{align}
    0 &\sim {\cal Z}_\mathrm{C}(0) 
      + e^{N_\mathrm{c} (\mu_\sigma + i \epsilon) /T} {\cal Z}_\mathrm{C}(N_\mathrm{c})
\nonumber\\
      &= A
      + e^{N_\mathrm{c} (\mu_\sigma + i\epsilon - M)/T} \, B_1,
\end{align}
and then
\begin{align}
    \mu_\sigma 
    &\sim M + \frac{T}{N_\mathrm{c}} \ln \Bigl( \frac{A}{B_1} \Bigr),
      \label{eq:mus}
\end{align}
where $M$ is the constituent quark mass which depends on the current quark mass and we here temporally introduce the infinitesimally small $\mu_\mathrm{I}=\epsilon$.
It should be noted that we here consider the sufficiently large but finite size system and thus phase transition is smeared.
It is, however, Lee-Yang zeros can appear very close to the real axis and thus we can evaluate the phase transition point from the finite size system without any inconsistencies.
In the large $N_\mathrm{c}$ with sufficiently low $T$, we can consider following three scenarios (I)-(III).
\begin{description}
   \item[Scenario (I)]
   If the second term is $1/N_\mathrm{c}$ suppressed in Eq.\,(\ref{eq:mus}) and the $\cos(N_\mathrm{c} \theta)$ function dominates the oscillating behavior of ${\cal Z}_\mathrm{GC}(\theta)$, the transition point is coincident with the value of the quarkyonic phase transition energy scale.
  This is corresponding to the situation that $A$ and $B_1$ are the same order.
  \item[Scenario (II)] If $\ln (A/B_1)$ in the second term is proportional to $N_\mathrm{c}$, the transition point is shifted from $M$.
  \item[Scenario (III)] If $\ln (A/B_1)$ in the second term is proportional to $N_\mathrm{c}^2$ or higher, the transition point is diverged. 
  At low $T$, this scenario seems as the unfeasible scenario because of the following discussion based on the physical degree of freedom.
\end{description}
Since the Stefan-Boltzmann limit of the pressure is proportional to $N_\mathrm{c}^2$ because of the gluon degree of freedom, but it should be lower number in the sufficiently low $T$ region because the physical degree of freedoms are gluballs and baryons even at finite $\theta$.
Therefore, in the present energy scale, the scenario (I) or (II) is realized because the pressure and the partition function are related with each other via Eq.\,(\ref{Eq:pressure});
for example, see Ref.\,\cite{Bringoltz:2005rr} for the normalized pressure on the lattice by that with the Stefan-Boltzmann limit for several $N_\mathrm{c}$.
These discussion, of course, must be modified with larger $T$.

Above discussions are implicitly assumed that condensates are homogeneous.
Even if we consider the spatial inhomogeneity, above discussions are valid if we replace $\sigma$ as $\sigma ({\bf x})$.
Then, we can discuss the inhomogensously chiral-symmetry broken phase which is the so called the real kink crystal~\cite{Buballa:2014tba} by introduce the external field which breaks the transnational invariance ($J_\mathrm{\delta}$) and take the $J_{\delta} \to 0$ limit.
In addition, via the same procedure as in the diquark condensate which will be explained in the next section, we can consider the condensation of the neutral pion $\langle \pi^0({\bf x}) \rangle (\mu_\mathrm{R})$ because the pion condensate does not appear in the finite $\theta$ region in the $J_{\delta} \to 0$ limit.
Then, we can consistently investigate the dual chiral density wave~\cite{Nakano:2004cd} which is another type of the inhomogeneous chiral-symmetry broken phase in addition to the real kink crystal.

\section{Color superconductivity}
\label{Sec:diquark}

Here, we discuss the color superconducting from the viewpoint of the canonical sectors.
However, the color superconductivity is $1/N_\mathrm{c}$ suppressed~\cite{Shuster:1999tn} and thus it is difficult to obtain clear picture from the present canonical sector approach without actual numerical calculations.
Therefore, we here show some very qualitative discussions on the color superconducting.

To discuss the color superconducting phase with the canonical ensemble method, we must introduce the external field ($J_\Delta$) and take the $J_\Delta \to 0$ limit even at finite $\theta$.
Of course, at least for CFL, we can consider the gauge-invariant (color singlet) order-parameter, but it is not possible for the 2SC because unbroken global symmetries in the 2SC phase are the same with those in the hadron phase; for example see Refs.\,\cite{Rajagopal:2000wf,Alford:2007xm}.
Therefore, we here consider the diquark condensate by introducing the gauge symmetry breaking external field.
With the external field, we have the following relation where $\Delta$ means the particular diquark operator;
\begin{align}
    \langle \Delta \rangle(\mu_\mathrm{R})
    &= \sum_{n=-\infty}^\infty
        e^{N_\mathrm{c} n \mu_\mathrm{R}/T}
       \frac{{\cal Z}_\mathrm{C}(N_\mathrm{c} n)}{{\cal Z}_\mathrm{GC}(\mu_\mathrm{R})}
       \langle \Delta \rangle_\mathrm{C} (N_\mathrm{c} n),
\end{align}
where $\langle \Delta \rangle_\mathrm{C}$ means the expectation value of $\Delta$ operator calculated from the $\theta$ region. 
We here assume that all flavors are degenerated.

Since we introduce $J_\mathrm{\Delta}$ and consider the $J_\Delta \to 0$ limit, we can assume
\begin{align}
    \langle \Delta \rangle_\mathrm{C} (N_\mathrm{c} n) = \epsilon_n,
\end{align}
where $\epsilon_n$ is infinitesimal value and approaches to $0$ in the $J_\Delta \to 0$ limit.
It should be noted that we may have finite $\langle \Delta \rangle_\mathrm{C}(N_\mathrm{c}n)$ even if we take $J_\Delta \to 0$ because we here consider the sufficiently large but finite size system. Therefore, more strictly speaking, we have $\langle \Delta \rangle_\mathrm{C}(N_\mathrm{c}n) \to \epsilon_\mathrm{V}$ where $\epsilon_\mathrm{V}$ is the infinitesimal value which approaches to $0$ with the $V \to \infty$ limit.
In below, we concentrate on the infinitesimal value induced by nonzero $J_\mathrm{\Delta}$ and thus we simply write $\epsilon_\mathrm{V}=0$ because following discussions are almost unchanged if we remain $\epsilon_\mathrm{V}$:
At least for discussions in this section, the thermodynamic limit is not the matter and thus one can consider following discussions are done with the thermodynamic limit.

As the same discussion in Sec.\,\ref{Sec:large_nc} and \ref{Sec:example}, we can expect the exponential suppression factor depends on $M_\mathrm{B}=N_\mathrm{c} M$ which can be factored out from the canonical partition function.
Therefore, we may write the condensate as
\begin{align}
    &\langle \Delta \rangle (\mu_\mathrm{R})
    \nonumber\\
    &= \sum_{n=-\infty}^{\infty}
        e^{N_\mathrm{c} n (\mu_\mathrm{R} - M)/T}
       \Bigl[ e^{N_\mathrm{c} n M/T} 
       \frac{{\cal Z}_\mathrm{C}(N_\mathrm{c} n)}{{\cal Z}_\mathrm{GC}(\mu_\mathrm{R})}
       \Bigr]\,
       \epsilon_n
    \nonumber\\
    &\sim \frac{1}{{\cal Z}_\mathrm{GC}(\mu_\mathrm{R})} \Bigl[
          f_0 \, \epsilon_0 
        + f_1 \, e^{N_\mathrm{c} (\mu_\mathrm{R} - M)/T} \epsilon_1 
        + \cdots \Bigr],
       \label{Eq:cond}
\end{align}
where $f_n$ are coefficients depend on $T$, $n$, $M$ and $J_\Delta$ at moderate $\mu_\mathrm{R}$, and the suppression factor is factored out from them; this means that we factor out the suppression factor from $\mathcal{Z}_\mathrm{C}(N_\mathrm{C}n)$ in the second line of Eq.\,(\ref{Eq:cond}).
It should be noted that we assume that $\exp(-n N_\mathrm{c} M/ T)$ factor appears for each canonical sector as in the last line of Eq.\,(\ref{Eq:cond}) because they are $nN_\mathrm{c}$-quarks contributions.
In the present work, we are working with $T \ll 1$ and thus exponential factors can lead the strong suppression at small $\mu_\mathrm{R}$.

Our interest is that above expression can lead the following situation;
\begin{align}
    \langle \Delta \rangle (\mu_\mathrm{R}) &=
    \begin{dcases}
    ~0 & (\mathrm{low} \, \mu_\mathrm{R})\\
    ~\mathrm{nonzero} & (\mathrm{moderate\, and\,  high}\,\mu_\mathrm{R})
    \end{dcases}
    .
    \label{Eq:cond_f}
\end{align}
In other wards, the present question is "Is it possible to manifest Eq.\,(\ref{Eq:cond_f}) in the canonical ensemble method?".
The answer of the question can be easily understood from the final expression in Eq.\,(\ref{Eq:cond}); if $\exp[N_\mathrm{c} (\mu_\mathrm{R} - M)/T]$ becomes $\epsilon^{-1}$-order, above situation (\ref{Eq:cond_f}) is realized.
Therefore, if the second term is the first nontrivial contribution to $\langle \Delta \rangle$, the necessary energy-scale $\Lambda_\Delta$ for the phase boundary of the color superconducting phase is about $\mu_\mathrm{R} \sim M$.
When $\mu_\mathrm{R}$ becomes larger value than $\Lambda_\Delta$, higher-order terms will join the game and thus this discussion will be modified.
Detailed value of $\Lambda_\Delta$ is strongly affected by the actual dependence of $f_n$ on $n$; since the higher oscillating modes are expected to be $1/n$ suppressed with increasing $n$ and thus the second term in Eq.\,(\ref{Eq:cond}) is expected to be the first nontrivial contribution for the nonzero diquark condensate at moderate $\mu_\mathrm{R}$. 
Therefore, the answer for our question "Is it possible to manifest Eq.\,(\ref{Eq:cond_f}) in the canonical ensemble method?" is yes at least with the present simplified setup.
This is the one of possible scenario for the color superconducting from the viewpoint of the canonical sectors.

To discuss one another possible scenario for the situation (\ref{Eq:cond_f}), we consider following setup.
To simplify our discussion, we here assume
\begin{align}
    e^{N_\mathrm{c} n M/T} \frac{{\cal Z}_\mathrm{C}(N_\mathrm{c} n)}{{\cal Z}_\mathrm{GC}(\mu_\mathrm{R})}
    \epsilon_n \sim f^n \epsilon,
\end{align}
where $f$ is a constant which is less than $1$ and $\epsilon$ denotes the infinitesimal value; $f^n$ mimics the $1/n$ suppression of higher-order oscillating modes.
With this assumption, Eq.\,(\ref{Eq:cond}) becomes 
the sum of infinite geometric series as
\begin{align}
    \langle \Delta \rangle (\mu_\mathrm{R})
    &\sim \frac{1}
               {1 - r}
               \epsilon,
\label{Eq:series}
\end{align}
with
\begin{align}
    r = f e^{N_\mathrm{c} (\mu_\mathrm{R} - M)/T}.
\end{align}
From the equation, we have the critical chemical potential ($\mu_\mathrm{\Delta}$) for the color superconducting as
\begin{align}
\mu_\mathrm{\Delta} = M - \frac{T}{N_\mathrm{c}} \ln f,
\end{align}
from $(1-r)=\epsilon$.
Since $f$ is smaller than $1$ in the present setting, $\mu_\mathrm{\Delta}$ is shifted to larger value comparing with $M$.
To more correct discussion, we need the exact expression of the sum; the coefficient $f$ should have $\mu_\mathrm{R}$-dependence via ${\cal Z}_\mathrm{GC}(\mu_\mathrm{R})$ and should remove the divergence. 
With increasing $\mu_\mathrm{R}$, the denominator of Eq.\,(\ref{Eq:series}) can become $\epsilon^1$-order around $\mu_\mathrm{R} \sim M$; its detailed value is affected by the actual value of $f$. 
Therefore, we can have the same result obtained in the previous paragraph in the present setup.
These means that the situation for the chiral condensate and the diquark condensate are different even if we base on the same canonical ensemble.

Above discussions are just theoretical expectation and thus it is not exact because we do not known actual behavior of the second term in Eq.\,(\ref{Eq:cond}) with $J_\Delta \to 0$ limit at low $T$.
It should be noted that each $\epsilon_n$ should depends on $1/N_\mathrm{c}$ and thus it becomes zero in the large $N_\mathrm{c}$ limit; it is consistent with the large $N_\mathrm{c}$ QCD picture.
These results can be checked when $J_\Delta$ is introduced to the NJL-type model in the canonical ensemble method with the multi-precision calculation used in Refs.\,\cite{Wakayama:2019hgz,Wakayama:2020dzz}.

\section{Summary}
\label{Eq:summary}

In this paper, we have discussed the confinement-deconfinement nature of QCD at moderate real chemical potential ($\mu_\mathrm{R}$) with low temperature ($T$) from the viewpoint of the canonical ensembles.
The canonical partition function (${\cal Z}_\mathrm{C}$) with fixed quark number $n$ is constructed by using the grand-canonical partition function (${\cal Z}_\mathrm{GC}$) at finite $\theta$ via the Fourier transformation and the fugacity expansion where $\theta:=\mu_\mathrm{I}/T$ with the imaginary chemical potential ($\mu_\mathrm{I}$).

We have assumed that ${\cal Z}_\mathrm{GC}(\theta)$ is consist of the constant part and the $2\pi/N_\mathrm{c}$-periodic oscillating mode which is proportional to $\cos (N_\mathrm{c}\theta)$; it can be justified at sufficiently low $T$.
With the matching with the energy scale of the quarkyonic phase transition known in the large $N_\mathrm{c}$ QCD, we can discuss the confinement-deconfinement nature from the change of relevant canonical sectors with varying $\mu_\mathrm{R}$. It should be noted that this behavior can be obtained in the QCD effective model with some ansatz.
There is the balanced region where two canonical sectors are equally contributes to ${\cal Z}_\mathrm{GC} (\mu_\mathrm{R})$, and it is very tiny in the large $N_\mathrm{c}$ because the region is $1/N_\mathrm{c}$ suppressed.

In the realistic QCD case, it is difficult to make the clear discussion about the quarkyonic phase, but we can still expect that the change of relevant canonical sectors with varying $\mu_\mathrm{R}$.
In this case, the balanced region between nearest-neighbour canonical-sectors is not strongly suppressed and thus there are continues change of the system.
This behavior may be consistent with the recent expectation at moderate $\mu_\mathrm{R}$ based on the quakyonic phase and the soft surface delocalization scenario.
Unfortunately, the numerical calculation to construct the canonical sectors is quite hard; we need the multi-precision calculation in the Fourier transformation~\cite{Fukuda:2015mva,Wakayama:2020dzz,Wakayama:2019hgz} and thus actual numerical calculation is our future work.

In addition to the quakyonic phase, we have presented some qualitative discussions about the color superconducting and the chiral symmetry restoration from the canonical sectors.
This paper guarantees that the canonical ensemble method can work at moderate $\mu_\mathrm{R} $ even with low $T$ and then the quakyonic and color superconducting phases and also the chiral symmetry restoration can be described in the method.
Since the exploration of the QCD phase diagram from the lattice QCD simulation with the canonical ensemble method at moderate $\mu_\mathrm{R}$ and low $T$ must be the long journey, we hope present results become the marker for the right direction.

\begin{acknowledgments}
This work is supported in part by the Grants-in-Aid for Scientific Research from JSPS (No. 18K03618, 19H01898 and 20K03974).
\end{acknowledgments}

\appendix

\bibliography{ref.bib}

\end{document}